**Regular Article**

# Cosmological constrains on new generalized Chaplygin gas model

Fataneh Salahedin[1], Reza Pazhouhesh[1,a], Mohammad Malekjani[2]

[1] Department of Physics, Faculty of Sciences, University of Birjand, Birjand, Iran
[2] Department of Physics, Bu-Ali Sina University, Hamedan 65178, Iran

**Abstract** We use different combinations of data samples to investigate the new generalized Chaplygin gas (NGCG) model in the context of dark energy (DE) cosmology. Using the available cosmological data, we put constraints on the the free parameters of NGCG model based on the statistical Markov chain Monte Carlo method. We then find the best fit values of cosmological parameters and those confidence regions in NGCG cosmology. Our result for the matter density parameter calculated in NGCG model is in excellent agreement with that of the standard $\Lambda$CDM cosmology. We also find that the equation of state of DE of the model slightly favors the phantom regime. We show that the big tension between the low- and high-redshift observations appearing in $\Lambda$CDM universe to predict the Hubble constant $H_0$ can be alleviated in NGCG model. However, from the statistical point of view, our results show that the standard $\Lambda$CDM model fits the observations better than the NGCG cosmology.

## 1 Introduction

The analysis of various observational data, including those of cosmic microwave background (CMB) [1,2], supernovae type Ia (SNIa) [3,4], large-scale structures (LSS), baryon acoustic oscillation (BAO) [5] and other cosmic observations strongly suggest that the present universe is undergoing an accelerated phase of expansion. In order to explain this acceleration, an unknown component with negative pressure the so-called dark energy (DE) was proposed [6]. The earliest and simplest model for DE is the cosmological standard $\Lambda$ [7] with constant equation of state (EoS) parameter equal to $-1$. This model can successfully describe observations, while it indeed encounters some of theoretical problems; for example the coincidence and fine-tuning problems [8,9]. Therefore, some of other DE models have been proposed in the literature, such as quintessence [10], phantom [11,12], holographic [13], oscillating Quintom [14], agegraphic [15] and Ghost [16]. Although these models can solve or alleviate the problems of $\Lambda$CDM, they should be confirmed by the cosmic observations. Many investigations have been done to examine these DE models in the light of observational data (reader can see [17–21]). Besides the DE models, modified gravity theories such as scalar tensor cosmology [22] and braneworld models [23] were proposed to solve the challenge of acceleration of universe.

---

[a] e-mail: rpazhouhesh@birjand.ac.ir (corresponding author)

The Chaplygin gas model is a candidate of DE that unifies dark matter and DE. In fact, this model plays a dual role at different epochs of the history of the universe. It plays the role of pressure-less dark matter in the early universe and DE at the late time. The other property of this model is that the Chaplygin gas model belongs to the category of dynamical DE with time-varying EoS parameter differs from −1 alleviating the coincidence problem in standard ΛCDM cosmology [24]. The simplest form of this model, the so-called standard Chaplygin gas (SCG), was proposed in the field of cosmology by Kamenshchik et al. [24] and Gorini et al. [25]. Although the SCG model explains the late time-accelerated expansion of universe, this model cannot interpret the scenario of the structure formation in the universe [26,27]. To solve this problem, the SCG model is generalized into the generalized Chaplygin gas (GCG) model [28–30]. Same as the SCG model, the GCG model can obtain the accelerated expansion of the universe [31]. This model has been widely studied in the literature and has confirmed by observations. The results of [31] show that the GCG model is well fitted by WMAP, CMBR and BAO data sets. They showed that one can assume the GCG model as an interacting form of ΛCDM [31]. Furthermore, a new version of generalized Chaplygin gas (NGCG) model, which can be a kind of interacting wCDM model, was proposed in [32]. They showed that the NGCG model has totally a dual role to interpret the interacting $w$CDM parametrization, where the interaction between DE and dark matter is characterized by a constant parameter. The authors of [32] have also performed a statefinder analysis on the NGCG model and found some discrimination between NGCG model and other DE scenarios. They also by performing the statistical likelihood analysis using different data of type Ia supernovae, CMB and LSS have provided a fairly tight constraints on the free parameters of the model.

Nowadays, there are different samples of observational data in a wide range of redshifts which can be used to put new updated constraints on parameters of cosmological models. In this work, we study the NGCG model using different data sets of cosmological observations. To do this, we use some available of these samples including type Ia supernovae (SNIa) from the Pantheon catalog [33] and the Union2.1 catalog [34], Big Bang nucleosynthesis (BBN) [35], BAO [36,37], CMB from the results of WMAP observations [38] and recently updated observational Hubble parameter data H (z) extracted from cosmic chronometers. These data can reveal the role of DE in the dynamics of the accelerated expansion of the universe. In our analysis, we use the Markov chain Monte Carlo (MCMC) method to constrain the cosmological parameters of the NGCG model. Finally, using the Akaike information criteria [39], we compare the NGCG model with standard ΛCDM cosmology. The summary of this paper is as follows. We start by introducing the NGCG model in Sect. 2. In Sect. 3, we present the observational constrains on the free parameters of NGCG model. In Sect. 4, we present our numerical results and finally in Sect. 5, we summarize our results and expose the main conclusions.

## 2 New generalized Chaplygin gas model

We briefly introduce the NGCG model in this section. Assuming that the universe is flat with the Friedman–Robertson–Walker (FRW) metric, the equation of state of NGCG fluid is given as follows [32]:

$$\rho_{\text{NGCG}} = -\frac{\tilde{A}(a)}{\rho_{\text{NGCG}}^{\alpha}}, \qquad (1)$$

where $\alpha$ is the constant parameter of NGCG fluid and $a$ is the scale factor. The NGCG fluid consists of dark matter and DE $\rho_{\text{de}} \sim a^{-3(1+w_d)}$ where $w_d$ is the EOS parameter. The energy

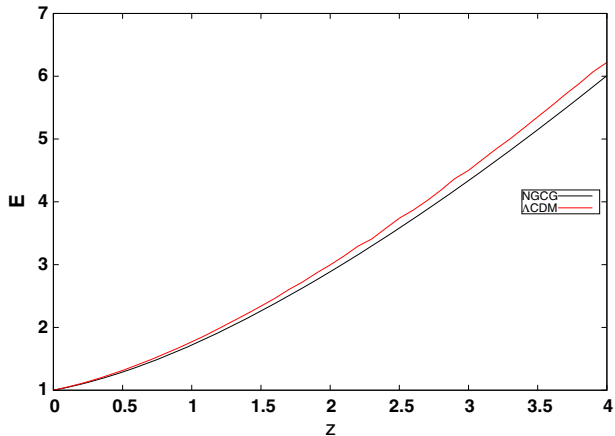

**Fig. 1** Redshift evolution of Hubble parameter E in terms of redshift parameterizations of NGCG and $\Lambda$CDM models

density of the NGCG can be written as [32]:

$$\rho_{NGCG} = \left[ Aa^{-3(1+w_d)(1+\alpha)} + Ba^{-3(1+\alpha)} \right]^{1/1+\alpha} \quad (2)$$

$$A + B = \rho_{NGCG0}^{1+\alpha} \quad (3)$$

where $A$ and $B$ are positive constants and the $\tilde{A}(a)$ defined as follows

$$\tilde{A}(a) = -w_d A a^{-3(1+w_d)(1+\alpha)} \quad (4)$$

Therefore, the NGCG energy density is defined as [32]:

$$\rho_{NGCG} = \rho_{NGCG0} a^3 \left[ 1 - A_S + A_S a^{-3w_d(1+\alpha)} \right]^{\frac{1}{1+\alpha}} \quad (5)$$

where $A_S$ is a parameter of NGCG fluid. Notice that based on the last equation, the constant parameters of NGCG are redefined as $A_s$ and $\alpha$. Therefore, the density of DE and dark matter in NGCG model can be written as follows

$$\rho_{de} = \rho_{de0} a^{-3[1+w_d(1+\alpha)]} \times \left[ 1 - A_s + A_s a^{-3w_d(1+\alpha)} \right]^{\frac{1}{1+\alpha}-1} \quad (6)$$

$$\rho_{dm} = \rho_{dm0} a^{-3} \left[ 1 - A_s + A_s a^{-3w_d(1+\alpha)} \right]^{\frac{1}{1+\alpha}-1} \quad (7)$$

We can easily see that the GCG model is retrieved when the EOS parameter of DE component, $w_d$, is equal to $-1$. Additionally, if $\alpha = 0$, this model reduces to wCDM model. It is worth mentioning that $\alpha$ describes the interaction between DE and dark matter.

When $\alpha > 0$, the energy is transferred from dark matter to DE. On the contrary, the energy is transferred from DE to dark matter in the case of $\alpha < 0$ [32]. The Hubble parameter in a flat universe filled by radiation, baryonic matter and NGCG fluid can be written as

$$H(a) = H_0 E(a), \quad (8)$$

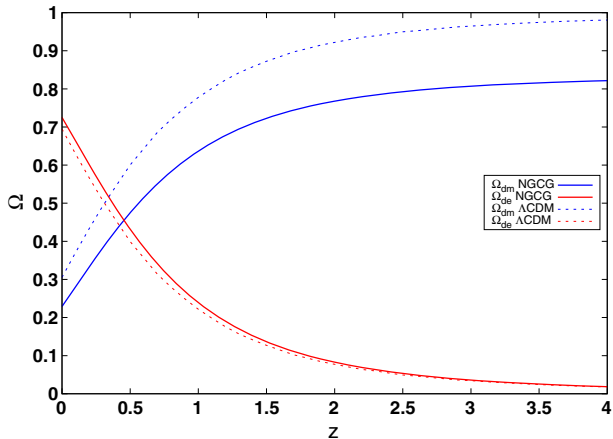

**Fig. 2** Evolution of the fractional energy density of nonrelativistic matter and dark energy components in terms of redshift parameterizations of NGCG and $\Lambda$CDM

where

$$E(a)^2 = (1 - \Omega_{b0} - \Omega_{r0}) a^{-3} \left[1 - A_s \left(1 - a^{-3w_d(1+\alpha)}\right)\right]^{\frac{1}{1+\alpha}} + \Omega_{b0} a^{-3} + \Omega_{r0} a^{-4}, \quad (9)$$

where $\Omega_{b0}$ and $\Omega_{r0}$ are the present values of dimensionless energy densities of baryonic matter and radiation, respectively, and $H_0$ is the Hubble constant. Solving Eq. (9), we show the redshift evolution of Hubble parameter $E$ for NGCG parametrization in Fig. 1. The same quantity is also plotted for concordance $\Lambda$CDM model. Here, we fix the cosmological parameters based on the best fit values obtained in second and third columns of Table 3. In Fig. 2, we show the evolution of fractional energy density of nonrelativistic dark matter and DE component in terms of redshift for NGCG and standard $\Lambda$CDM models. We see that similarly to $\Lambda$CDM model, the NGCG model can interpret the matter dominated universe at early times and DE domination at later times.

## 3 Observational constraints on NGCG model parameters

In this section, we use the well-known and available cosmological data sets to put constraints on the free parameters of the NGCG model. Given a cosmological model with a set of free parameters and using a set of observational data points, we can define a merit function to quantify the agreement between the model and observations. In this regard, by maximizing the degree of agreement, one can obtain the best fit values of the free parameters of the model. Moreover, our analysis should provide the error bar of each parameter reasonable measure of the goodness of the fit. Notice that if the model cannot fit the observations, then the obtained best fit values of the free parameters are obviously meaningless. Here, we use the minimum chi-squared ($\chi^2$) technique for model fitting procedure. Given a set of data points $D$ and a cosmological model, $M(x, \mathbf{p})$, where vector $\mathbf{p}$ includes the free parameters of the model, the

$\chi^2$ function is defined as follows

$$\chi^2 = \sum_i \frac{1}{\sigma_i^2}[D_i - M(x_i \mid \mathbf{p})]^2, \qquad (10)$$

where $\sigma_i$ is the error of data point $i$. The best fit values of the free parameters $\mathbf{p}$ are obtained by minimizing the $\chi^2$ function. Notice that the above equation for calculating $\chi^2$ function is valid when the observational data points are not correlated. If we use correlated data points, we should use the new formula

$$\chi^2 = \sum_{i,j}[D_i - M(x_i \mid \mathbf{p})]\mathcal{Q}_{ij}[D_j - M(x_j \mid \mathbf{p})]. \qquad (11)$$

where $\mathcal{Q}_{ij}$ is the inverse of the covariance matrix. The covariance matrix explains the covariance between observational data points.

Notice that when we compute different $\chi^2$ functions for different data sets, we should sum all of the $\chi_n^2$ functions. So finding the best fit values of free parameters requires the minimizing of the sum of $\chi_n^2$ functions. We allow the free parameters of the model walk in a wide range of values. To do this, we perform a Markov chain Monte Carlo (MCMC) analysis. In this work, we compute the $\chi^2$ function for different SNIa, CMB, BAO, BBN and H(z) data sets.

$$\chi_{\text{tot}}^2(\mathbf{p}) = \chi_{\text{SN}}^2 + \chi_{\text{BAO}}^2 + \chi_{\text{CMB}}^2 + \chi_{\text{BBN}}^2 + \chi_H^2, \qquad (12)$$

The first data set we have considered is the distance modulus of SNIa including 580 distinct data points from Union2.1 sample [34] and 1048 distinct data points from Pantheon catalog catalog [40]. Finally, we have a short comparison between these two samples. The $\chi^2$ function for SNIa samples is therefore given by

$$\chi_{\text{sn}}^2 = \sum_i \frac{[\mu_{\text{th}}(z_i) - \mu_{\text{ob}}(z_i)]^2}{\sigma_i^2} \qquad (13)$$

where $\mu_{\text{th}}(z_i)$ is the theoretical value of distance modulus at the specific redshift $z_i$ and $\mu_{\text{obs}}$ is the corresponding observational value. Here, $\sigma_i$ is the error bar of the observational data. For a given cosmological model, we have $\mu_{\text{th}}(z) = 5\log_{10}\left[(1+z)\int_0^z \frac{dz}{E(z)}\right] + \mu_0$, where $\mu_0 = 42.384 - 5\log_{10} h$ [41]. The second data set which we consider is the BAO sample. The BAO includes six distinct measurements of the baryon acoustic scale. These data points and their references are summarized in Table 1. Since the BAO data are correlated, the $\chi_{\text{BAO}}^2$ function is given by [42]

$$\chi_{\text{BAO}}^2 = Y^T C_{\text{BAO}}^{-1} Y, \qquad (14)$$

where the vector $Y$ is given by

$$Y = d(0.1) - d_1, \frac{1}{d(0.35)} - \frac{1}{d_2}, \frac{1}{d(0.57)} - \frac{1}{d_3},$$
$$d(0.44) - d_4, d(0.6) - d_5, d(0.73) - d_6.$$

The quantity $d(z_i)$ is defined as the ratio of comoving sound horizon at the baryon drag epoch, $r_s$, divided into a function of angular diameter distance, $D_V(z)$, as

$$d(z) = \frac{r_s(z_{\text{drag}})}{D_v(z)}, \qquad (15)$$

**Table 1** Current available data points for BAO measurements which we use in our analysis

| $z$ | $d_i$ | References |
| --- | --- | --- |
| 0.106 | 0.336 | [43] |
| 0.35 | 0.113 | [44] |
| 0.57 | 0.073 | [45] |
| 0.44 | 0.0916 | [46] |
| 0.6 | 0.0726 | [46] |
| 0.43 | 0.0592 | [46] |

where $r_s$ is given by

$$r_s(a) = \int_0^a \frac{c_s \, da}{a^2 H(a)} \tag{16}$$

In the above equations, $c_s$ is the baryon sound speed. Finally, the function of angular diameter distance $D_v(z)$ is defined by

$$D_v(z) = \left[ (1+z)^2 D_A^2(z) \frac{z}{H(z)} \right]^{\frac{1}{3}} \tag{17}$$

where $D_A$ is the angular diameter distance. We refer the reader to [42], to see the fitting formula for estimated redshift of baryon drag epoch, $z_{\text{drag}}$, which we adopt it to compute $d(z_i)$ in our work. The baryon sound speed in Eq. (16) is given by

$$c_s(a) = \frac{1}{\sqrt{3\left(1 + \frac{3\Omega_{b0}}{\Omega_{\gamma 0}} a\right)}} \tag{18}$$

where $\Omega_{\gamma 0} = 0.469 \times 10^{-5} \, h^{-2}$ [42]. We also adopt the covariance matrix $C_{\text{BAO}}^{-1}$ given in [42] as

$$C_{\text{BAO}}^{-1} = \begin{bmatrix} 4444.4 & 0 & 0 & 0 & 0 & 0 \\ 0 & 34.602 & 0 & 0 & 0 & 0 \\ 0 & 0 & 206611 & 0 & 0 & 0 \\ 0 & 0 & 0 & 24532.1 & -25137.7 & 12099.1 \\ 0 & 0 & 0 & -25137.7 & 134598.4 & -64783.9 \\ 0 & 0 & 0 & 12099.1 & -64783.9 & 128837.6 \end{bmatrix} \tag{19}$$

The third sample which we use in our numerical analysis is the data of the position of t CMB acoustic peak. This sample is very important to provide resonable constrain on DE models. The position of this peak is given by $(l_a, R, z_*)$, where $z_*$ is the recombination epoch, $R$ is the scale distance to recombination epoch and

$$l_a = \pi \frac{D_A(z_*)}{r_s(z_*)}. \tag{20}$$

Also, the prior distance $R$ is given by

$$R = \sqrt{\Omega_{m0}} H_0 D_A(z_*). \tag{21}$$

We also adopt the fitted formula for the epoch for recombination presented in [42]. We use the WMAP data set for the position of acoustic pick of CMB measurements [47]. Notice

**Table 2** H(z) data points including their references

| $z$ | H(z) | $\sigma_H$ | References |
|---|---|---|---|
| 0 | 73.24 | 1.74 | [47] |
| 0.07 | 69 | 19.6 | [48] |
| 0.1 | 69 | 12 | [49] |
| 0.12 | 68.6 | 26.2 | [48] |
| 0.17 | 83 | 8 | [49] |
| 0.1791 | 75 | 4 | [50] |
| 0.1993 | 75 | 5 | [50] |
| 0.2 | 72.9 | 29.6 | [48] |
| 0.27 | 77 | 14 | [49] |
| 0.28 | 88.8 | 36.6 | [48] |
| 0.35 | 82.7 | 8.4 | [51] |
| 0.3519 | 83 | 14 | [50] |
| 0.4 | 95 | 17 | [49] |
| 0.48 | 97 | 62 | [49] |
| 0.5929 | 104 | 13 | [50] |
| 0.6797 | 92 | 8 | [50] |
| 0.7812 | 105 | 12 | [50] |
| 0.8754 | 125 | 17 | [50] |
| 0.88 | 90 | 40 | [49] |
| 0.9 | 177 | 23 | [49] |
| 1.037 | 154 | 20 | [50] |
| 1.3 | 168 | 17 | [49] |
| 1.43 | 177 | 18 | [49] |
| 1.53 | 140 | 14 | [49] |
| 1.75 | 202 | 40 | [49] |
| 2.33 | 244 | 8 | [52] |

that the data points are correlated. Then, we have [53]

$$\chi^2_{\text{CMB}} = X^T_{\text{CMB}} C^{-1}_{\text{CMB}} X_{\text{CMB}}, \qquad (22)$$

where

$$X_{\text{CMB}} = \begin{pmatrix} l_a - 302.40 \\ R - 1.7264 \\ z_* - 1090.88 \end{pmatrix}, \qquad (23)$$

and

$$C^{-1}_{\text{CMB}} = \begin{pmatrix} 3.182 & 18.253 & -1.429 \\ 18.253 & 11887.879 & -193.808 \\ -1.429 & -193.808 & 4.556 \end{pmatrix}. \qquad (24)$$

The forth data set used in our analysis is the Hubble data points extracted from cosmic chronometers. These data and their references are collected in Table 2. The $\chi^2$ function for non-correlated Hubble data is given by

$$\chi_H^2 = \sum_i \frac{[H_{\text{th}}(z_i) - H_{\text{obs}}(z_i)]^2}{\sigma_i^2} \quad (25)$$

where $H_{\text{th}}$ is the theoretical value of Hubble parameter, $H_{\text{obs}}$ is the observational value and the parameters $\sigma_i$ is the corresponding uncertainty. The fifth data set is the Big Bang Nucleosynthesis (BBN) measurements which provide a data point for energy density of baryons $\Omega_{b0}$. The $\chi_{\text{BBN}}^2$ is simply given by

$$\chi_{\text{BBN}}^2 = \left(\frac{(\Omega_{\text{b0}}h^2 - 0.022)}{0.002}\right)^2 \quad (26)$$

By jointing all data samples, we perform a likelihood analysis based on the MCMC algorithm to calculate the minimum of $\chi_{\text{tot}}^2$ and the best fit values of the cosmological parameters. Our results and discussion are presented in next section.

## 4 Results and discussion

In this study, two samples of type Ia supernovae (SNIa) data, i.e., Pantheon and Union2.1 catalogs, were used. Notice that the Pantheon sample contains more observational data at higher redshift which basically leads to tighter constrain rather than Union 2.1 sample. We consider two different jointed samples. Firstly, we joint the SNIa (Union 2.1) with BAO, BBN, CMB and Hubble data and secondly, we joint the SNIa (Pantheon) with with BAO, BBN, CMB and Hubble data. For both cases, we do our analysis and obtain the best fit values of cosmological parameters leading to finding the minimum of $\chi^2$ function. The numerical results are presented in Table 3. Also the $1\sigma$, $2\sigma$ and $3\sigma$ confidence contours for various cosmological parameters of $\Lambda$CDM ,and NGCG models are shown in Figs. 3, 4 and 5. We note that for $\Lambda$CDM model, we only use the Union 2.1 sample. We observe that the best fit value of EoS parameter of NGCG model restricts $w_\Lambda = -1$ even in $1\sigma$ contour. Notice that the best fit value prefers the phantom regime of EoS parameter, but the deviation from $w_\Lambda = -1$ cannot deviate from $1\sigma$ error. Also the best fit value of the energy density of nonrelativistic matter ($\Omega_{dm0} + \Omega_{b0}$) obtained in our analysis for NGCG model (both cases) is consistent with that of the $\Lambda$CDM cosmology even in $2\sigma$ uncertainty (95% confidence level). We know that in the context of observational cosmology, there is a big tension between the low-redshift observations and high-redshift CMB data in predicting the value of Hubble constant. Quantitatively speaking, the high-redshift CMB data predict $H_0 = 67.4^{+0.5}_{-0.5}$ km s$^{-1}$ Mpc$^{-1}$ [54], while from the Cepheid-calibrated SnIa at low redshifts we have $H_0 = 74.03^{+1.42}_{-1.42}$ km s$^{-1}$ Mpc$^{-1}$ [55]. The combination of all high- and low-redshift cosmological data can alleviate this tension causing to get lower values of $H_0$ closer to CMB predictions. In the case of $\Lambda$CDM cosmology (Union 2.1 sample of Supernovae used), our results for the combinations of all data sets show that $H_0 = 71.3^{+1.1,+2.2,+3.0}_{-1.1,-2.2,-2.9}$ (third column of Table 3), while in the case of the NGCG model (first column of Table 3) we have $H_0 = 70.41^{+0.92,+1.8,+2.3}_{-0.92,-1.7,-2.3}$. Hence, we see that in the case of NGCG model the tension is alleviated approximately around $1\sigma$ error. Our numerical results for NGCG model (both cases) show that the best fit value of interaction parameter $\alpha$ is negative at least in $1\sigma$ level. Hence ,we can say that the NGCG model prefers the phantom regime in which the energy transfers from DE to dark matter. As one can see in Table 3, in the case of jointed data sets:

**Table 3** Results of statistical likelihood analysis using the different combinations of cosmological data as (different SNIa catalog + CMB + BAO + BBN + H(z)) for NGCG model and standard ΛCDM universe

| | $H(z)$ + BAO + BBN + CMB + SNIa (Union2.1) | $H(z)$ + BAO + BBN + CMB + SNIa (pantheon) | ΛCDM model |
|---|---|---|---|
| $\Omega_b{}_0$ | $0.0457\,^{+0.0017}_{-0.0017}\,^{+0.0033}_{-0.0032}\,^{+0.0043}_{-0.0042}$ | $0.0460\,^{+0.0017}_{-0.0017}\,^{+0.0034}_{-0.0034}\,^{+0.0045}_{-0.0039}$ | $0.0446\,^{+0.0014}_{-0.0014}\,^{+0.0028}_{-0.0026}\,^{+0.0036}_{-0.0034}$ |
| $\eta = 1+\alpha$ | $0.981\,^{+0.018}_{-0.018}\,^{+0.033}_{-0.032}\,^{+0.043}_{-0.042}$ | $0.9443\,^{+0.0097}_{-0.0097}\,^{+0.020}_{-0.019}\,^{+0.026}_{-0.025}$ | – |
| $\Omega_{dm0}$ | $0.2353\,^{+0.0097}_{-0.0092}\,^{+0.019}_{-0.016}\,^{+0.025}_{-0.020}$ | $0.2508\,^{+0.0081}_{-0.0097}\,^{+0.017}_{-0.02}\,^{+0.023}_{-0.026}$ | $0.2229\,^{+0.0099}_{-0.0099}\,^{+0.020}_{-0.019}\,^{+0.027}_{-0.025}$ |
| $H_0$ | $70.41\,^{+0.92}_{-0.92}\,^{+1.8}_{-1.7}\,^{+2.3}_{-2.3}$ | $70.15\,^{+0.84}_{-0.84}\,^{+1.6}_{-1.6}\,^{+2.0}_{-2.3}$ | $71.3\,^{+1.1}_{-1.1}\,^{+2.2}_{-2.2}\,^{+3.0}_{-2.9}$ |
| $w_d$ | $-1.021\,^{+0.055}_{-0.055}\,^{+0.11}_{-0.11}\,^{+0.14}_{-0.15}$ | $-1.041\,^{+0.045}_{-0.045}\,^{+0.088}_{-0.091}\,^{+0.11}_{-0.12}$ | $-1$ |
| $A_s$ | $0.753\,^{+0.010}_{-0.010}\,^{+0.020}_{-0.020}\,^{+0.026}_{-0.027}$ | $0.7371\,^{+0.0097}_{-0.0086}\,^{+0.017}_{-0.019}\,^{+0.022}_{-0.024}$ | – |
| $\chi^2_{\min}$ | 591.4 | 1065.2 | 572.0 |

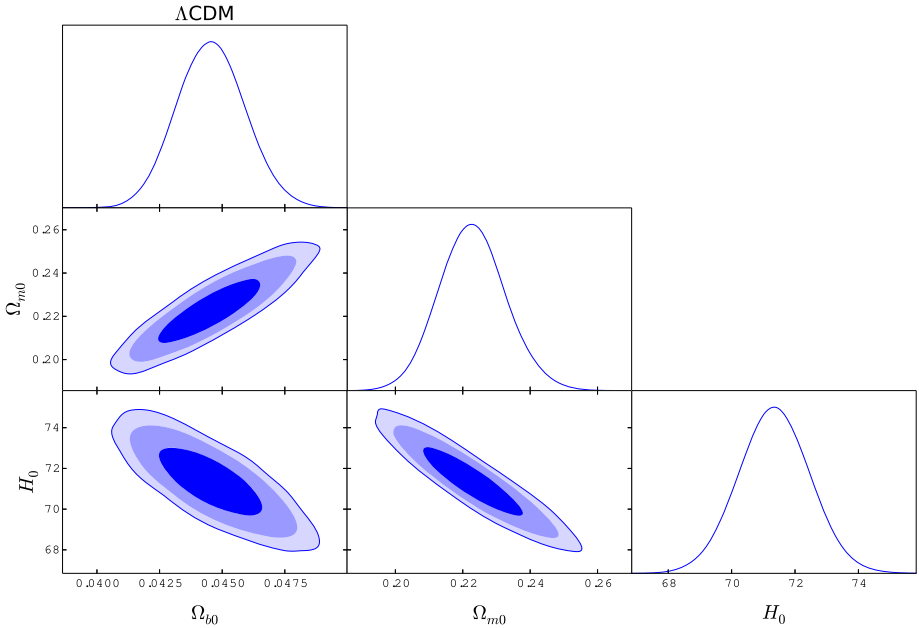

**Fig. 3** $1\sigma$, $2\sigma$ and $3\sigma$ contours for various cosmological parameters using the combined cosmological data SNIa (Union2.1 catalog) + CMB + BAO + BBN + $H(z)$ for $\Lambda$CDM model

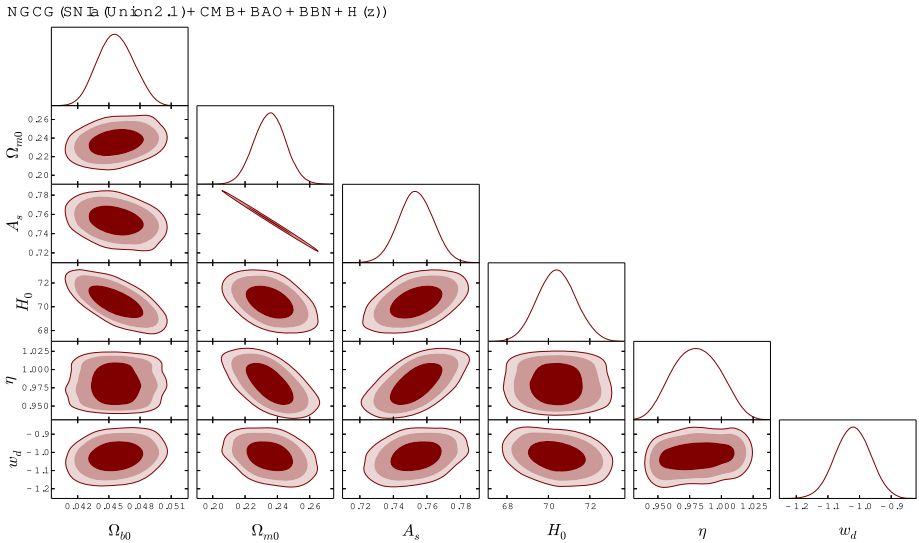

**Fig. 4** $1\sigma$, $2\sigma$ and $3\sigma$ contours for various cosmological parameters using the cosmological data SNIa (Union2.1 catalog) + CMB + BAO + BBN + $H(z)$ for NGCG model

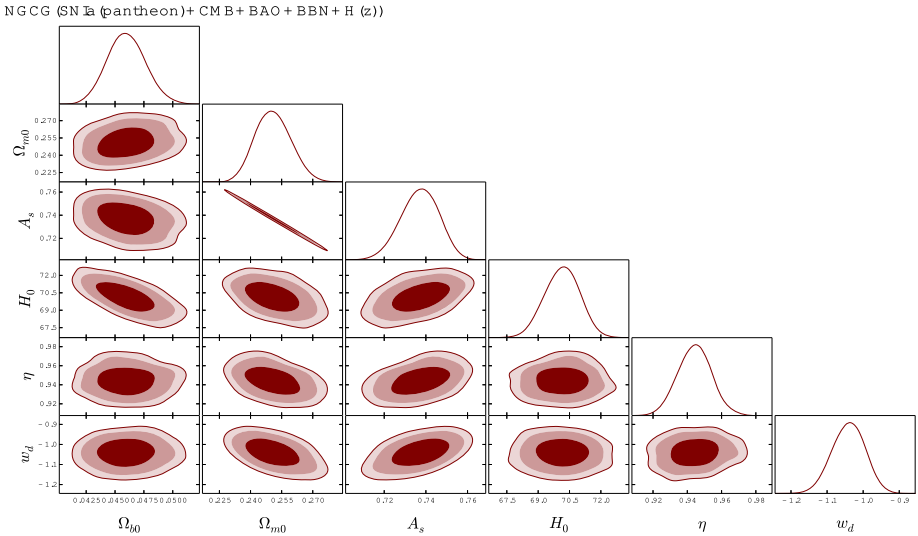

**Fig. 5** $1\sigma$, $2\sigma$ and $3\sigma$ contours for various cosmological parameters using the cosmological data of SNIa (Pantheon catalog) + CMB + BAO + BBN + $H(z)$ for NGCG model

$H(z)$ + BAO + CMB+ BBN + SNIa (Union2.1), we obtained $\chi^2_{\min} = 591.4$ for NGCG model, while this value for $\Lambda$CDM is $\chi^2_{\min} = 572.0$. Therefore, we can say that upon this combination of data points, $\Lambda$CDM model is fitted to observations better than the NGCG model. Furthermore, we should consider this point that the value of 591.4 for NGCG model is obtained in the presence of six free parameters, while in our analysis $\Lambda$CDM just has three free parameters. It is easy to know that more number of free parameters leads to improvement in fitting procedure. To remove the effect of extra parameters, we can use the well-known Akaike information criteria [39], AIC $= \chi^2_{\min} + 2k$ where $k$ is the number of free parameters. If we compute the AIC value for the models under study, we will obtain AIC$_{\Lambda\text{CDM}} = 578.0$ and AIC$_{\text{NGCG}} = 603.4$ which the big value of $\Delta$AIC $= 25.4$ indicates that the standard $\Lambda$CDM cosmology is in better agreement with the whole of cosmological observations. (For more information about the relevance of $\Delta$AIC for support to a given model, we refer the readers to [56].) This result shows that from the statistical likelihood analysis, the simple $\Lambda$CDM model is still the best model when we consider all the observational data. Notice that in previous studies regarding the dynamical DE, the same results have been obtained (for example, see [17–19,21,57]). On the base of our knowledge, there is still no dynamical DE with time- varying EoS parameter with AIC value lower than the standard $\Lambda$CDM model.

In next step of our analysis, we consider another combination of data samples to see that how our results depend on the including or excluding of Hubble data and Supernova data. To do this, we first exclude the Hubble data (first and second column of Table 4) and then exclude the Supernova samples (third column of Table 4). The $1\sigma$, $2\sigma$ and $3\sigma$ confidence regions are shown in Figs. 6, 7 and 8. One can easily see that in the presence of Supernova data, and the EoS parameter of NGCG model is in agreement with constant value $w_\Lambda = -1$ in $1\sigma$ contour. These results are independent of the Hubble data because by comparing the first and second columns of Tables 3 and 4, we observe the negligible differences. While if we exclude the Supernova data (third column of Table 4), the results are reasonably changed. We observe that, in this case, the EoS parameter differs from constant $w_\Lambda = -1$ approximately with $1\sigma$ distance, confirming the phantom like of NGCG model.

**Table 4** Results of statistical likelihood analysis using different combination of data sets as (BAO + CMB + BBN + SNIa (Union2.1), BAO + CMB + BBN + SNIa (pantheon), BAO + CMB + BBN + SNIa (pantheon) and H(z) + BAO + CMB + BBN)

| | BAO + CMB + BBN + SNIa (Union2.1) | BAO + CMB + BBN + SNIa (pantheon) | BAO + CMB + BBN + SNIa (pantheon) and H(z) + BAO + CMB + BBN + H(z) |
|---|---|---|---|
| $\Omega_{b0}$ | $0.0466 \; {}^{+0.0018 \; + 0.0039 \; + 0.0057}_{-0.0021 \; - 0.0036 \; - 0.0047}$ | $0.0469 \; {}^{+0.0018 \; + 0.0037 \; + 0.0050}_{-0.0018 \; - 0.0034 \; - 0.0042}$ | $0.0442 \; {}^{+0.0023 \; + 0.0046 \; + 0.0059}_{-0.0023 \; - 0.0043 \; - 0.0053}$ |
| $\eta = 1 + \alpha$ | $0.9796 \; {}^{+0.0085 \; + 0.017 \; + 0.019}_{-0.085 \; - 0.017 \; - 0.021}$ | $0.9528 \; {}^{+0.0074 \; + 0.017 \; + 0.023}_{-0.0074 \; - 0.016 \; - 0.020}$ | $0.982 \; {}^{+0.016 \; + 0.033 \; + 0.035}_{-0.016 \; - 0.034 \; - 0.035}$ |
| $\Omega_{dm0}$ | $0.2399 \; {}^{+0.0095 \; + 0.018 \; + 0.023}_{-0.0095 \; - 0.019 \; - 0.023}$ | $0.2512 \; {}^{+0.0089 \; + 0.017 \; + 0.024}_{-0.0089 \; - 0.016 \; - 0.021}$ | $0.230 \; {}^{+0.011 \; + 0.021 \; + 0.027}_{-0.011 \; - 0.019 \; - 0.024}$ |
| $H_0$ | $69.7 \; {}^{+1.1 \; + 2.2 \; + 3.2}_{-1.1 \; - 2.2 \; - 3.8}$ | $69.50 \; {}^{+0.99 \; + 1.9 \; - 2.5}_{-0.99 \; - 2.0 \; - 2.6}$ | $71.6 \; {}^{+1.6 \; + 3.1 \; + 3.9}_{-1.6 \; - 3.0 \; - 3.9}$ |
| $w_d$ | $-1.011 \; {}^{+0.056 \; + 0.10 \; + 0.13}_{-0.056 \; - 0.11 \; - 0.15}$ | $-1.036 \; {}^{+0.044 \; + 0.085 \; + 0.10}_{-0.044 \; - 0.087 \; - 0.12}$ | $-1.12 \; {}^{+0.13 \; + 0.23 \; + 0.30}_{-0.12 \; - 0.25 \; - 0.32}$ |
| $A_s$ | $0.748 \; {}^{+0.010 \; + 0.021 \; + 0.025}_{-0.010 \; - 0.020 \; - 0.025}$ | $0.7364 \; {}^{+0.0095 \; + 0.017 \; + 0.023}_{-0.0095 \; - 0.019 \; - 0.026}$ | $0.759 \; {}^{+0.011 \; + 0.021 \; + 0.026}_{-0.011 \; - 0.023 \; - 0.029}$ |
| $\chi^2_{min}$ | 573.1 | 1046.4 | 28.1 |

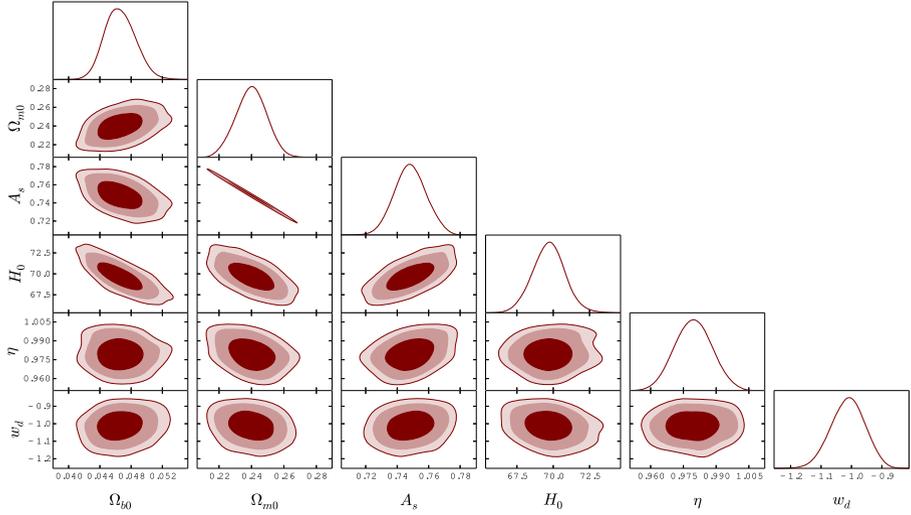

**Fig. 6** $1\sigma$, $2\sigma$ and $3\sigma$ contours for various cosmological parameters using the cosmological data of SNIa (Union2.1 catalog) + CMB + BAO + BBN for NGCG model

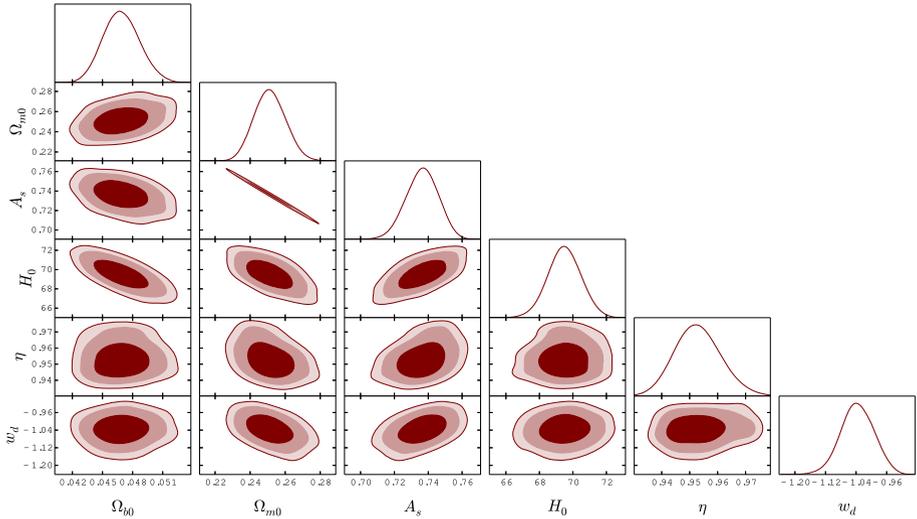

**Fig. 7** $1\sigma$, $2\sigma$ and $3\sigma$ contours for various cosmological parameters using the cosmological data of SNIa (Pantheon catalog) + CMB + BAO + BBN for NGCG model

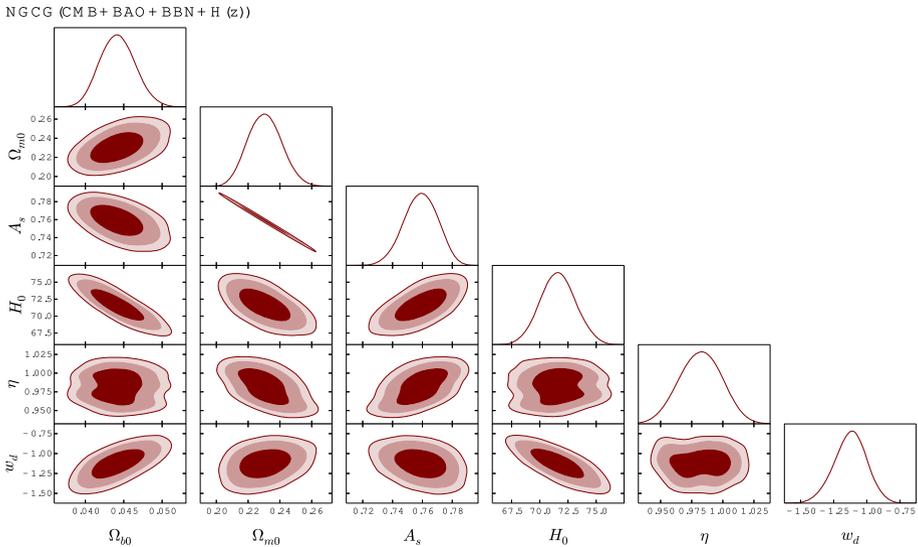

**Fig. 8** $1\sigma$, $2\sigma$ and $3\sigma$ contours for various cosmological parameters using the cosmological data CMB + BAO + BBN + $H(z)$ for NGCG model

## 5 Conclusions

We studied the cosmological properties of NGCG DE model and compared the results in the light of different combinations of observational data points. We also compare the results of NGCG model with those of $\Lambda$CDM to find the best model which can describe the evolution of cosmic fluid. Using different SNIa catalogs, we found that all combinations of data samples which we considered for NGCG model lead to the same value of matter energy density with that of the $\Lambda$CDM cosmology at least within $1\sigma$ uncertainty. Our results showed that the NGCG model can alleviate the tension between low-redshift observations with the Plank inferred value of $H_0$ in the amount of $1\sigma$ level better than the $\Lambda$CDM cosmology. Our results also showed that when we use one of the SNIa catalogs in our data combination, the value of $w_d$ comes closer to $-1$ while without SNIa catalogs, $w_d$ moves in to the phantom region beyond $1\sigma$ uncertainty. Comparing the values of $\chi^2_{\min}$ and AIC values which we have obtained for the models under study by using the same data combination, we found that the $\Lambda$CDM is still the best model to fit the all observational data in cosmology. Thus, we can conclude that the NGCG model can alleviate the coincidence problem and $H_0$ tension appearing in $\Lambda$CDM cosmology, but from the statistical likelihood analysis, the standard $\Lambda$CDM cosmology is still the best model to fit the whole of cosmological data in expanding universe.

## References

1. E. Komatsu, J. Dunkley, M.R. Nolta et al., ApJS **180**, 330 (2009)
2. E. Komatsu, K.M. Smith, J. Dunkley et al., ApJS **192**, 18 (2011)
3. A.G. Riess, A.V. Filippenko, P. Challis et al., AJ **116**, 1009 (1998)
4. S. Perlmutter, G. Aldering, G. Goldhaber et al., ApJ **517**, 565 (1999)
5. W.J. Percival, B.A. Reid, D.J. Eisenstein et al., MNRAS **401**, 2148 (2010)
6. P.A.R. Ade et al., A&A **594**, A20 (2016)

7. S. Weinberg, Mod. Phys. Rev. **61**, 527 (1989)
8. T. Padmanabhan, Phys. Rep. **380**, 235 (2003)
9. E.J. Copeland, M. Sami, S. Tsujikawa, Int. J. Mod. Phys. D **15**, 1753 (2006)
10. P.G. Debendetti, E.H. Stanley, Supercooled and glassy water. Phys. Today **56**(3), 40–46 (2003)
11. R.R. Caldwell, M. Kamionkowski, N.N. Weinberg, Phys. Rev. Lett. **91**, 071301 (2003). arXiv:astro-ph/0302506
12. M.R. Setare, Eur. Phys. J. C **50**, 991 (2007)
13. Q. Wu, Y.G. Gong, A.Z. Wang, J.S. Alcanizd, Phys. Lett. B **659**, 34 (2008)
14. B. Feng, M. Li, Y.-S. Piao, X. Zhang, Phys. Lett. B **634**, 101 (2006)
15. R.G. Cai, Phys. Lett. B **657**, 228 (2007)
16. R. Schutzhold, Phys. Rev. Lett. **89**, 081302 (2002)
17. A. Mehrabi, S. Basilakos, F. Pace, Mon. Not. R. Astron. Soc. **452**, 2930 (2015)
18. M. Malekjani, S. Basilakos, Z. Davari, A. Mehrabi, M. Rezaei, Mon. Not. R. Astron. Soc. **464**, 1192 (2017)
19. M. Rezaei, M. Malekjani, S. Basilakos, A. Mehrabi, D.F. Mota, Astrophys. J. **843**, 65 (2017)
20. M. Rezaei, Mon. Not. R. Astron. Soc. **485**, 4841 (2019)
21. M. Rezaei, M. Malekjani, J. Sola, Phys. Rev. D **100**, 023539 (2019)
22. B. Boisseau, G. Esposito-Farese, D. Polarski, A.A. Starobinsky, Phys. Rev. Lett. **85**, 2236 (2000)
23. G. Dvali, G. Gabadadze, M. Porati, Phys. Lett. B **485**, 208 (2000)
24. A.Y. Kamenshchik et al., Phys. Lett. B **511**, 265 (2001)
25. V. Gorini, et al., (2004). arXiv:gr-qc/0403062v2
26. H.B. Sandvik et al., Phys. Rev. D **69**, 123524 (2004)
27. R. Bean et al., Phys. Rev. D **68**, 023515 (2003)
28. A.Y. Kamenshchik, U. Moschella, V. Pasquier, Phys. Lett. B **511**, 265 (2001)
29. M.C. Bento, O. Bertolami, A.A. Sen, Phys. Rev. D **66**, 043507 (2002). arXiv:gr-qc/0202064
30. J.C. Fabris et al., Gen. Rel. Grav. **36**, 211 (2004)
31. T. Barreiro, O. Bertolami, P. Torres, Phys. Rev. D **78**, 043530 (2008)
32. X. Zhang, F.Q. Wu, J. Zhang, JCAP **0601**, 003 (2006)
33. D.M. Scolnic et al., Astrophys. J. **859**, 101 (2018)
34. N. Suzuki, D. Rubin, C. Lidman, G. Aldering et al., ApJ **746**, 85 (2012)
35. R.J. Cooke, M. Pettini, C.C. Steidel, Astrophys. J. **855**, 102 (2018)
36. D.J. Eisenstein et al., Astrophys. J. **633**, 560 (2005)
37. C. Blake, E. Kazin, F. Beutler, T. Davis, D. Parkinson et al., MNRAS **418**, 1707 (2011)
38. J. Dunkley et al., ApJS **180**, 306 (2009)
39. H. Akaike, ITAC **19**, 716 (1974)
40. D.M. Scolnic et al., Astrophys. J. **859**, 101 (2018)
41. A. Mehrabi, S. Basilakos, F. Pace, MNRAS **452**, 2930 (2015)
42. G. Hinshaw et al., ApJS **208**, 19 (2013)
43. F. Beutler, C. Blake, M. Colless, D.H. Jones, L. Staveley-Smith et al., MNRAS **416**, 3017 (2011)
44. N. Padmanabhan, X. Xu, D.J. Eisenstein, R. Scalzo, A.J. Cuesta et al., MNRAS **427**, 2132 (2012)
45. L. Anderson, E. Aubourg, S. Bailey, D. Bizyaev, M. Blanton et al., MNRAS **427**, 3435 (2013)
46. C. Blake, S. Brough, M. Colless, C. Contreras, W. Couch et al., MNRAS **415**, 2876 (2011)
47. A.G. Riess, S. Casertano, W. Yuan, L. Macri, J. An-derson, J.W. MacKenty, J.B. Bowers, K.I. Clubb, A.V. Filippenko, D.O. Jones, B.E. Tucker, ApJ **855**, 136 (2018)
48. C. Zhang, H. Zhang, S. Yuan et al., Res. Astron. Astrophys. **14**, 1221 (2014)
49. D. Stern, R. Jimenez, L. Verde et al., J. Cosmol. Astropart. Phys. **1002**, 008 (2010)
50. M. Moresco et al., J. Cosmol. Astropart. Phys. **1208006**, 11 (2012)
51. C.H. Chuang, Y. Wang, Mon. Not. R. Astron. Soc. **435**, 255 (2013)
52. J.E. Bautista et al., Astron. Astrophys. **603**, A12 (2017)
53. W. Hu, N. Sugiyama, ApJ **471**, 542 (1996)
54. N. Aghanim et al. (Planck) (2018). arXiv:1807.06209
55. A.G. Riess, S. Casertano, W. Yuan, L.M. Macri, D. Scolnic, Astrophys. J. **876**, 85 (2019)
56. K.P. Burnham, D.R. Anderson, *Model Selection and Multi-model Inference* (Springer, New York, 2002)
57. M. Malekjani, M. Rezaei, I.A. Akhlaghi, Phys. Rev. D **98**, 063533 (2018)
58. E. Komatsu et al., Astrophys. J. Suppl. **192**, 18 (2011)